\documentclass[twocolumn]{aastex631}

\usepackage{amsmath,amsthm,amssymb,amsfonts,xspace}

\shorttitle{Machine learning dark matter substructure with astrometric lensing}
\shortauthors{Mishra-Sharma}

\newcommand{\package}[1]{\textsl{#1}\xspace}
\newcommand{\eg}{{e.\,g.}\xspace}
\newcommand{\ie}{{i.\,e.}\xspace}

\newcommand{\healpix}{\package{HEALPix}}
\newcommand{\deepsphere}{\package{DeepSphere}}

\begin{document}\sloppy\sloppypar\raggedbottom\frenchspacing

\title{Inferring dark matter substructure with astrometric lensing beyond the power spectrum}

\author[0000-0001-9088-7845]{Siddharth Mishra-Sharma}
\affiliation{The NSF AI Institute for Artificial Intelligence and Fundamental Interactions}
\affiliation{Center for Theoretical Physics, Massachusetts Institute of Technology, Cambridge, MA 02139, USA}
\affiliation{Department of Physics, Massachusetts Institute of Technology, Cambridge, MA 02139, USA}
\affiliation{Department of Physics, Harvard University, Cambridge, MA 02138, USA}
\affiliation{Center for Cosmology and Particle Physics, Department of Physics, New York University, New York, NY 10003, USA}

\begin{abstract}\noindent
Astrometry---the precise measurement of positions and motions of celestial objects---has emerged as a promising avenue for characterizing the dark matter population in our Galaxy. By leveraging recent advances in simulation-based inference and neural network architectures, we introduce a novel method to search for global dark matter-induced gravitational lensing signatures in astrometric datasets. Our method based on neural likelihood-ratio estimation shows significantly enhanced sensitivity to a cold dark matter population and more favorable scaling with measurement noise compared to existing approaches based on two-point correlation statistics. We demonstrate the real-world viability of our method by showing it to be robust to non-trivial modeled as well as unmodeled noise features expected in astrometric measurements. This establishes machine learning as a powerful tool for characterizing dark matter using astrometric data. 
\end{abstract}


\section{Introduction and background}
\label{sec:intro}

Although there exists plenty of evidence for dark matter (DM) on galactic scales and above (see \citet{Green:2021jrr} for a recent overview), the distribution of DM clumps---subhalos---on sub-galactic scales is less well-understood and remains an active area of cosmological study. This distribution additionally correlates with and may provide clues about the underlying particle physics nature of dark matter (see \eg, \citet{Schutz:2020jox,Bode:2000gq,Dalcanton:2000hn}), highlighting its relevance across multiple domains.

\begin{figure*}[!htbp]
\centering
\includegraphics[width=0.98\textwidth]{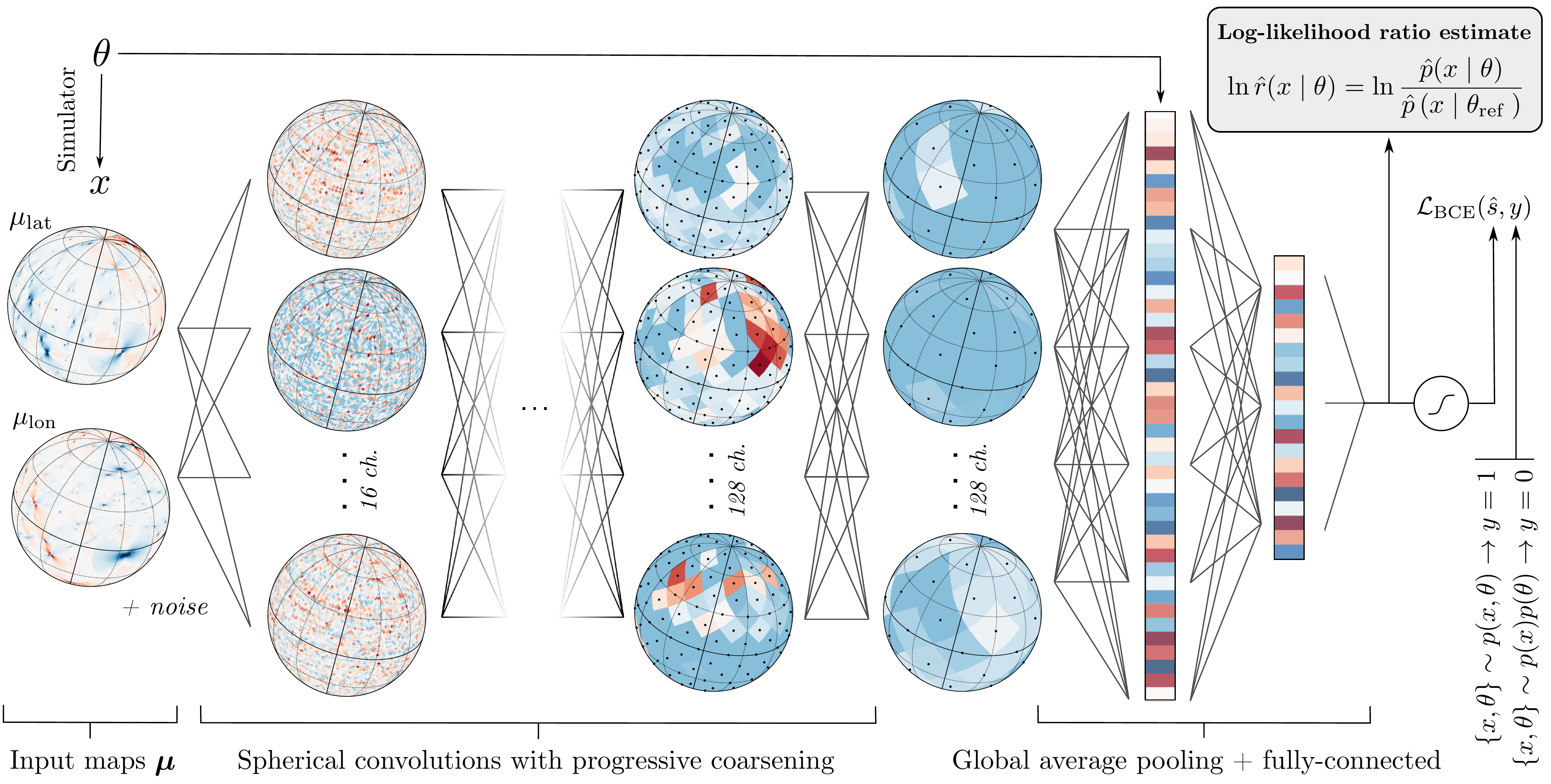}
\caption{A schematic illustration of the method and neural network architecture used in this work.}
\label{fig:model}
\end{figure*}

While more massive dark matter subhalos can be detected and studied through their association with luminous tracers such as bound stellar populations, subhalos with smaller masses $\lesssim 10^9\,\mathrm M_\odot$ are not generally associated with luminous matter~\citep{Fitts:2016usl,2017MNRAS.467.2019R}, rendering their characterization challenging. Gravitational effects provide one of the few avenues to probe the distribution of these otherwise-invisible subhalos~\citep{Buckley:2017ijx}. Gravitational lensing \ie, the bending of light from a background source due to a foreground mass, is one such effect and has been proposed in various incarnations as a probe of dark subhalos. 
Strong gravitational lensing, for example, has been used to infer the presence of dark matter substructure in galaxies outside of our own~\citep{Hezaveh:2016ltk,Vegetti:2009cz,Gilman:2019nap,Vegetti:2012mc}.
Astrometric lensing, on the other hand, has recently emerged as a promising way to characterize the dark matter subhalo population within the Milky Way.

Astrometry refers to the precise measurement of the positions and motions of luminous celestial objects like stars and galaxies. Gravitational lensing of these background objects by a moving foreground mass, such as a dark matter subhalo, can imprint a characteristic, correlated signal on their measured kinematics (angular velocities and/or accelerations). \citet{VanTilburg:2018ykj} introduced several methods for extracting this signature, including computing convolutions of the expected lensing signal on astrometric datasets and detecting local kinematic outliers. \citet{Mondino:2020rkn} applied the former method to data from the \emph{Gaia} satellite, obtaining constraints on the abundance of dark compact objects in the Milky Way and showcasing the applicability of astrometric dark matter searches in a practical setting. Finally, \citet{Mishra-Sharma:2020ynk} proposed using the angular power spectrum of the astrometric field as an observable to infer the population properties of subhalos in our Galaxy, leveraging the collective, correlated signal of a large subhalo sample. 

Astrometric datasets are inherently high-dimensional, consisting of positions and kinematics of potentially millions of objects. Especially when the expected signal consists of the collective imprint of a large number of lenses, characterizing their population properties involves marginalizing over all possible configurations of subhalos, rendering the likelihood intractable and usually necessitating the use of simplified data representations like the power spectrum. While effective, such simplification can result in loss of information compared to that contained in the original dataset when the expected signal is non-Gaussian in nature. The existence of systematic effects that are degenerate with a putative signal in the low-dimensional summary domain can further inhibit sensitivity. 

The dawn of the era of precision astrometry, with the \emph{Gaia} satellite~\citep{2016A&A...595A...1G} having recently delivered the most precise astrometric dataset to-date~\citep{2018A&A...616A...1G,2018A&A...616A...2L,2021A&A...649A...1G} and surveys including the Square Kilometer Array (SKA)~\citep{Fomalont:2004hr,Jarvis:2015tqa} and Roman Space Telescope~\citep{2019JATIS...5d4005W,2013arXiv1305.5425S,2019arXiv190205569A} set to achieve further leaps in sensitivity over the next decade, calls for methods that can extract more information from these datasets than is possible using existing techniques. In this direction, \citet{Vattis:2020kaa} proposed using a binary classifier in order to detect either the presence or absence of a substructure signal in astrometric maps. In this paper, we introduce an \emph{inference} approach that uses spherical convolutional neural networks---exploiting the symmetry structure of the signal and data domain---in conjunction with parameterized classifiers~\citep{Cranmer:2015bka,Baldi:2016fzo} in order to estimate likelihood ratios associated with the abundance of a cold dark matter population directly from a binned map of the astrometric velocity field. 
We show that our method outperforms established proposals based on the two-point correlation statistics of the astrometric field, both in absolute sensitivity as well as its scaling with measurement noise. 
{While we focus on the specific domain application to astrometric dark matter searches, we note that the method as presented here is broadly applicable to data sampled on the celestial sphere, which is ubiquitous in astrophysics and cosmology. More generally, the paper showcases how neural network architectures suited to processing real-world data structures---in our cases, pixelated vector fields of velocities on the celestial sphere---can be combined with advancements in simulation-based inference in order to directly perform inference on complex, high-dimensional datasets without resorting to the use of simplified summary statistics.}

The remainder of this paper is organized as follows. Section~\ref{sec:model} describes the forward model (Sec.~\ref{sec:forward_model}), the established approach based on the two-point statistics of the astrometric field (Sec.~\ref{sec:power_spectrum}), and various components of the method introduced in this paper (Secs.~\ref{sec:likelihood_ratio}--\ref{sec:training}). In Sec.~\ref{sec:experiments} we present results on simulated data, with the baseline results and diagnostic tests shown in Sec.~\ref{sec:baseline}. In Secs.~\ref{sec:lowell_noise} and \ref{sec:dr2_noise} we assess the robustness of our method to sources of unmodeled as well as modeled features expected in real-world applications. We conclude in Sec.~\ref{sec:conclusions}.

\section{Model and inference}
\label{sec:model}

\subsection{The forward model}
\label{sec:forward_model}

Our datasets consist of the 2-dimensional angular velocity map of background sources on the celestial sphere. {In order to define the forward model we need to specify the properties of background sources as well as the population properties of dark matter subhalos acting as gravitational lenses. We focus in this work on subhalos within the canonical cold dark matter scenario; details of the population model along with a the prescription for computing the induced velocity signal are provided in App.~\ref{app:forward_model_details}.} The subhalo fraction {$f_\mathrm{sub}\in \mathbb R$}, quantifying the expected fraction of the mass of the Milky Way contributed by subhalos in the range $10^{-6}$--$10^{10}\,\mathrm{M}_\odot$, is taken to be the parameter of interest.

We take our source population to consist of remote, point-like galaxies known as quasars which, due to their large distances from the Earth, are not expected to have significant intrinsic angular velocities. We assume the sources to be isotropically-distributed in the baseline configuration, {and further study the effect of relaxing this assumption using an existing catalog of quasars from \emph{Gaia}'s second data release (DR2)}. The velocity maps are assumed to be spatially binned, and we use  a \healpix binning~\citep{Gorski:2004by} with resolution parameter \texttt{nside=64}, corresponding to $N_\mathrm{pix}$ = 49,152 pixels over the full sky with pixel area $\sim 0.8\,\mathrm{deg}^2$. The values within each pixel then quantify the average latitudinal and longitudinal velocity components of quasars within that pixel. An example of the induced velocity signal on part of the celestial sphere, projected along the Galactic latitudinal and longitudinal directions and exhibiting dipole-like structures, is shown in the leftmost column of Fig.~\ref{fig:model}. {This pixelization level was motivated by the results of~\citet{Mishra-Sharma:2020ynk}, which showed the typical angular size of cold dark matter subhalos significantly contributing to the astrometric lensing signal to be much larger than the degree-scale pixel size used here. This can also be seen from the simulated signal realizations in Fig.~\ref{fig:model}}

In order to enable a comparison with traditional approaches---which are generally not expected to be sensitive to a cold dark matter subhalo population with next-generation astrometric surveys~\citep{VanTilburg:2018ykj,Mishra-Sharma:2020ynk}---we benchmark using an optimistic observational configuration corresponding to measuring the proper motions of $N_q = 10^8$ quasars {over the fully sky} with noise $\sigma_{\mu} = 0.1\,\mu\mathrm{as}\,\mathrm{yr}^{-1}$. {The final input maps $x\in\mathbb R^{2\,N_\mathrm{pix}}$ are obtained by combining the simulated signal with a realization of the noise model.}

\subsection{The power spectrum approach} 
\label{sec:power_spectrum}

\citet{Mishra-Sharma:2020ynk} introduced an approach for extracting the astrometric signal due to a dark matter subhalo population by decomposing the observed map into its angular (vector) power spectrum. The power spectrum is a summary statistic ubiquitous in astrophysics and cosmology and quantifies the amount of correlation contained at different spatial scales. In the case of data on a sphere, the basis of spherical harmonics is often used, and the power spectrum then encodes the correlation structure on different multipoles $\ell$. The power spectrum effectively captures the linear component of the signal and, when the underlying signal is a Gaussian random field, captures \emph{all} of the relevant information contained in the map(s)~\citep{Tegmark:1996qt}.
The expected signal in the power spectrum domain can be evaluated semi-analytically using the formalism described in \citet{Mishra-Sharma:2020ynk} and, assuming a Gaussian likelihood, the expected sensitivity can be computed using a Fisher forecasting approach. We use this prescription as a comparison point to the method introduced here.

While effective, reduction of the full astrometric map to its power spectrum results in loss of information; this can be seen from the fact that the signal in the leftmost column of Fig.~\ref{fig:model} is far from Gaussian. Furthermore, the existence of correlations on large angular scales due to \eg, biases in calibration of celestial reference frames~\citep{2018A&A...616A..14G} or systematic variations in measurements taken over different regions of the sky introduces degeneracies with a putative signal and precludes their usage in the present context. For this reason multipoles $\ell < 10$ were discarded in \citet{Mishra-Sharma:2020ynk}, degrading the projected sensitivity.

\subsection{Likelihood-ratio estimation using parameterized classifiers} 
\label{sec:likelihood_ratio}

Recent advances in machine learning have enabled methods that can be used to efficiently perform inference on models defined through complex simulations; see \citet{Cranmer:2019eaq} for a recent review. Here, we make use of neural likelihood-ratio estimation~\citep{Cranmer:2015bka,Baldi:2016fzo,Brehmer:2018eca,Brehmer:2018hga,Brehmer:2018kdj,Hermans:2019ioj}, previously applied to the problem of inferring dark matter substructure using observations of strong gravitational lenses~\citep{Brehmer:2019jyt} and cold stellar streams~\citep{Hermans:2020skz}. 

Given a classifier that can distinguish between samples $\{x\} \sim p(x\mid\theta)$ drawn from parameter points $\theta$ and those from a fixed reference hypothesis $\{x\} \sim p(x\mid\theta_\mathrm{ref})$, the decision function output by the optimal classifier $s(x, \theta) = {p(x\mid\theta)}/{\left(p(x\mid\theta) + p(x\mid\theta_\mathrm{ref})\right)}$ is one-to-one with the likelihood ratio, $r(x\mid \theta) \equiv {p(x\mid\theta)}/{p(x\mid\theta_\mathrm{ref})}  = {s(x, \theta)}/{\left(1 - s(x, \theta)\right)}$, a fact appreciated as the likelihood-ratio trick~\citep{Cranmer:2015bka,mohamed2017learning}. 
The classifier $s(x, \theta)$ in this case is a neural network that can work directly on the high-dimensional data, and is parameterized by the parameter of interest $\theta$ by having it included as an additional input feature. In order to improve numerical stability and reduce dependence on the fixed reference hypothesis $\theta_\mathrm{ref}$, we follow \citet{Hermans:2019ioj} and train a classifier to distinguish between data-sample pairs from the joint distribution $\{x, \theta\} \sim p(x,\theta)$ and those from a product of marginal distributions $\{x, \theta\} \sim p(x)p(\theta)$ (defining the reference hypothesis and in practice obtained by shuffling samples within a batch) using the binary cross-entropy (BCE) loss as the optimization objective. 

{We briefly highlight the advantages of this method over traditional paradigms for simulation-based inference such as Approximate Bayesian Computation (ABC,~\citet{10.1214/aos/1176346785,sisson2018handbook}). In ABC, samples $\{x\}$ from the forward model are compared to a particular dataset $x'$, with the approximate posterior defined through the set of parameters whose corresponding samples most closely match the dataset of interest according to a similarity metric. In our case, the curse of dimensionality would require a manual reduction of the raw datasets $x\in\mathbb R^{2\,N_\mathrm{pix}}$ into lower-dimensional summaries $f(x)\in\mathbb R^{n}$ with $n \ll 2\,N_\mathrm{pix}$ (\eg, the power spectrum), in order to enable tractable inference. A similarity metric and tolerance threshold $||f(x)-f(x')|| < \epsilon$ must additionally be specified in order to trade off between sample efficiency and inference precision. The machine learning-based method, on the other hand, uses neural networks in order to directly extract useful representations from high-dimensional datasets and can learn a continuous mapping from the data to the statistic of interest, in our case the likelihood ratio. Finally, ABC inference has to be performed anew for each dataset of interest. Our method, in contrast, is \emph{amortized}---following an upfront computational cost associated with training the likelihood-ratio estimator, evaluation on a new sample can be performed almost instantaneously. This allows us to efficiently test our model and compute diagnostics such as statistical coverage over large data samples.}

\subsection{Extracting information from high-dimensional astrometric maps} 
\label{sec:neural_network}

Since our data consists of a velocity field sampled on a sphere, we use a spherical convolutional neural network in order to directly learn useful representations from these maps that are efficiently suited for the downstream classification task. Specifically, we make use of \deepsphere~\citep{2020arXiv201215000D,deepsphere_cosmo}, a graph-based convolutional neural network tailored to data sampled on a sphere. For this purpose, the \healpix grid can be cast as a weighted undirected graph with $N_\mathrm{pix}$ vertices and edges connecting each pixel vertex to its set of 8 neighboring pixels. The weighted adjacency matrix over neighboring pixels $(i, j)$ is given by $A_{ij} = \exp \left(-{\Delta r_{ij}^{2}}/{\rho^{2}}\right)$ where $\Delta r_{ij}$ specifies the 3-dimensional Euclidean distance between the pixel centers and the widths $\rho$ are obtained from \citet{2020arXiv201215000D}. \deepsphere then efficiently performs convolutions in the spectral domain using a basis of Chebychev polynomials as convolutional kernels~\citep{2016arXiv160609375D}; here, we set $K=4$ as the maximum polynomial order. 

\begin{figure*}[!htbp]
\centering
\includegraphics[width=0.95\textwidth]{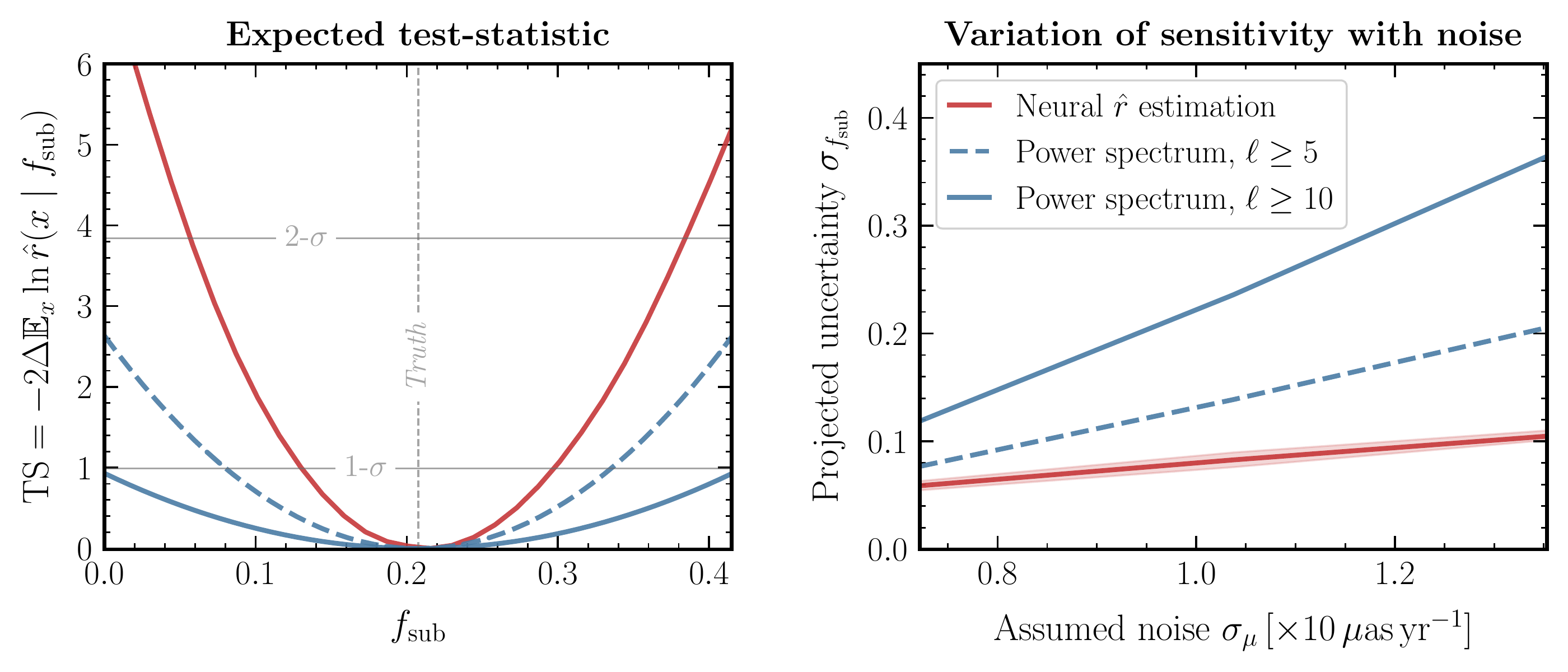}
\caption{\emph{(Left)} The expected log-likelihood ratio test-statistic (TS) profile for a cold dark matter population as a function of substructure fraction $f_\mathrm{sub}$ obtained using the neural likelihood-ratio estimation method introduced in this work (solid red line) compared with the corresponding profiles for existing approaches using power spectrum summaries with different multipole thresholds $\ell \gtrsim 5$ (dashed blue line) and $\ell \gtrsim 10$ (solid blue line). The vertical dotted line indicates the true benchmark value of the parameter $f_\mathrm{sub}$ in the test dataset. Our method shows enhanced sensitivity to a cold dark matter population compared to traditional approaches. \emph{(Right)} Scaling of the expected sensitivities, quantified by the respective 1-$\sigma$ uncertainties, with per-object instrumental noise. For the machine learning-based approach, the band quantifies the middle-95\% containment of the inferred 1-$\sigma$ uncertainty. Our method shows a more favorable scaling with assumed measurement noise.}
\label{fig:experiment}
\end{figure*}

All inputs are normalized to zero mean and unit standard deviation across the training sample. Starting with 2 scalar input channels representing the two orthogonal (Galactic latitude and longitude) components of the velocity vector map, we perform a graph convolution operation, increasing the channel dimension to 16 followed by a batch normalization, ReLU nonlinearity, and downsampling the representation by a factor of 4 with max pooling into the next coarser \healpix resolution. Pooling leverages scale separation, preserving important characteristics of the signal across different resolutions. 
Four more such layers are employed, increasing the channel dimension by a factor of 2 at each step until a maximum of 128, with maps after the last convolutional layer having resolution \texttt{nside=2} corresponding to 48 pixels. At this stage, we average over the spatial dimension (known as global average pooling~\citep{lin2014network}) in order to encourage approximate rotation invariance, outputting 128 features onto which the parameter of interest {$f_\mathrm{sub}\in\mathbb R$} is appended. These features are passed through a fully-connected network with (1024, 128) hidden units and ReLU activations outputting the classifier decision $\hat s$ by applying a sigmoidal projection.

{We note that the signal in our case does not respect strict rotation invariance---as described in App.~\ref{app:forward_model_details}, the motion of the Sun relative to the frame of rest of the Milky Way induces a preferred direction in the velocities of dark matter subhalos relative to our frame of reference, breaking the rotation symmetry of the signal. The anisotropy in the data domain is further exacerbated in the case of a realistic noise model (as explored in Sec.~\ref{sec:dr2_noise} below) where measured uncertainties vary along different directions on the celestial sphere. We finally note that by representing the input angular velocity vector field in terms of two independent scalar channels, we explicitly break the rotation equivariance of spherical convolutions due to differences in how scalar and vector representations transform under rotations (see, \eg, \citet{DBLP:conf/nips/EstevesMD20}). Given these limitations, we leave a detailed study of the equivariance properties desired of the architecture in the context of our application to future work.}

\subsection{Model training and evaluation} 
\label{sec:training}

$10^5$ maps from the forward model were produced, with 15\% of these held out for validation. {Samples containing between 0 and 300 subhalos in expectation over the mass range $10^{8}$--$10^{10}\,\mathrm{M}_\odot$, approximately corresponding to substructure fractions $f_\mathrm{sub}$ between 0 and 0.4, were generated from a uniform proposal distribution.} The estimator was trained using a batch size of 64 for up to 50 epochs with early stopping if the validation loss had not improved after 10 epochs. The \textsc{Adam} optimizer~\citep{kingma2017adam} was used with initial learning rate $10^{-3}$ decayed through cosine annealing. A coarse grid search was used to inform the architecture and hyperparameter choices in this work. {Experiments were performed on NVIDIA RTX8000 GPUs, taking $\sim10$ minutes per training epoch for a total training time of $\sim6$--9 hours contingent on early stopping.}

For a given test map, the log-likelihood ratio profile can be obtained by evaluating the trained estimator for different values of $f_\mathrm{sub}$ while keeping the input map fixed. The network output prior to the final sigmoidal projection directly gives the required log-likelihood ratio estimate: $\ln\hat r = S^{-1}(\hat s)$, where $S$ is the sigmoid function~\citep{Hermans:2019ioj,Hermans:2020skz}.
Figure~\ref{fig:model} presents an illustrative summary of the neural network architecture and method used in this work. \vspace{0.5cm}

\section{Experiments on simulated data}
\label{sec:experiments}

\subsection{Baseline results and diagnostics}
\label{sec:baseline}

We evaluate our trained likelihood-ratio estimator on maps drawn from a benchmark configuration motivated by \citet{Hutten:2016jko,Springel:2008cc}, containing 150 subhalos in expectation between $10^{8}$--$10^{10}\,\mathrm{M}_\odot$ and corresponding to $f_\mathrm{sub} \simeq 0.2$. The left panel of Fig.~\ref{fig:experiment} shows the expected log-likelihood ratio test-statistic (TS) as a function of substructure fraction $f_\mathrm{sub}$ for this nominal configuration. This is obtained by evaluating the trained estimator on 100 test maps over a uniform grid in $f_\mathrm{sub}$ and taking the point-wise mean. Corresponding curves using the power spectrum approach are shown in blue, using minimum multipoles of $\ell \geq 5$ (dashed) and $\ell \geq 10$ (solid). Thresholds corresponding to 1- and 2-$\sigma$ significance assuming a $\chi^2$-distributed TS are shown as the horizontal grey lines. We see that sensitivity gains of over a factor of $\sim 2$ can be expected for this particular benchmark when using the machine learning approach compared to the traditional power spectrum approach. No significant bias on the central value of the inferred DM abundance relative to the overall uncertainty scale is observed.

The right panel of Fig.~\ref{fig:experiment} shows the scaling of expected 1-$\sigma$ uncertainty on substructure fraction $f_\mathrm{sub}$ with assumed noise per quasar, keeping the number of quasars fixed (red, with the line showing the median and shaded band corresponding to the middle-95\% containment of the uncertainty inferred over 50 test datasets) compared to the power spectrum approach (blue lines). A far more favorable scaling of the machine learning approach is seen compared to the power spectrum approach, suggesting that it is especially advantageous in low signal-to-noise regimes that are generally most relevant for dark matter searches.

\begin{figure}[!htbp]
\centering
\includegraphics[width=0.46\textwidth]{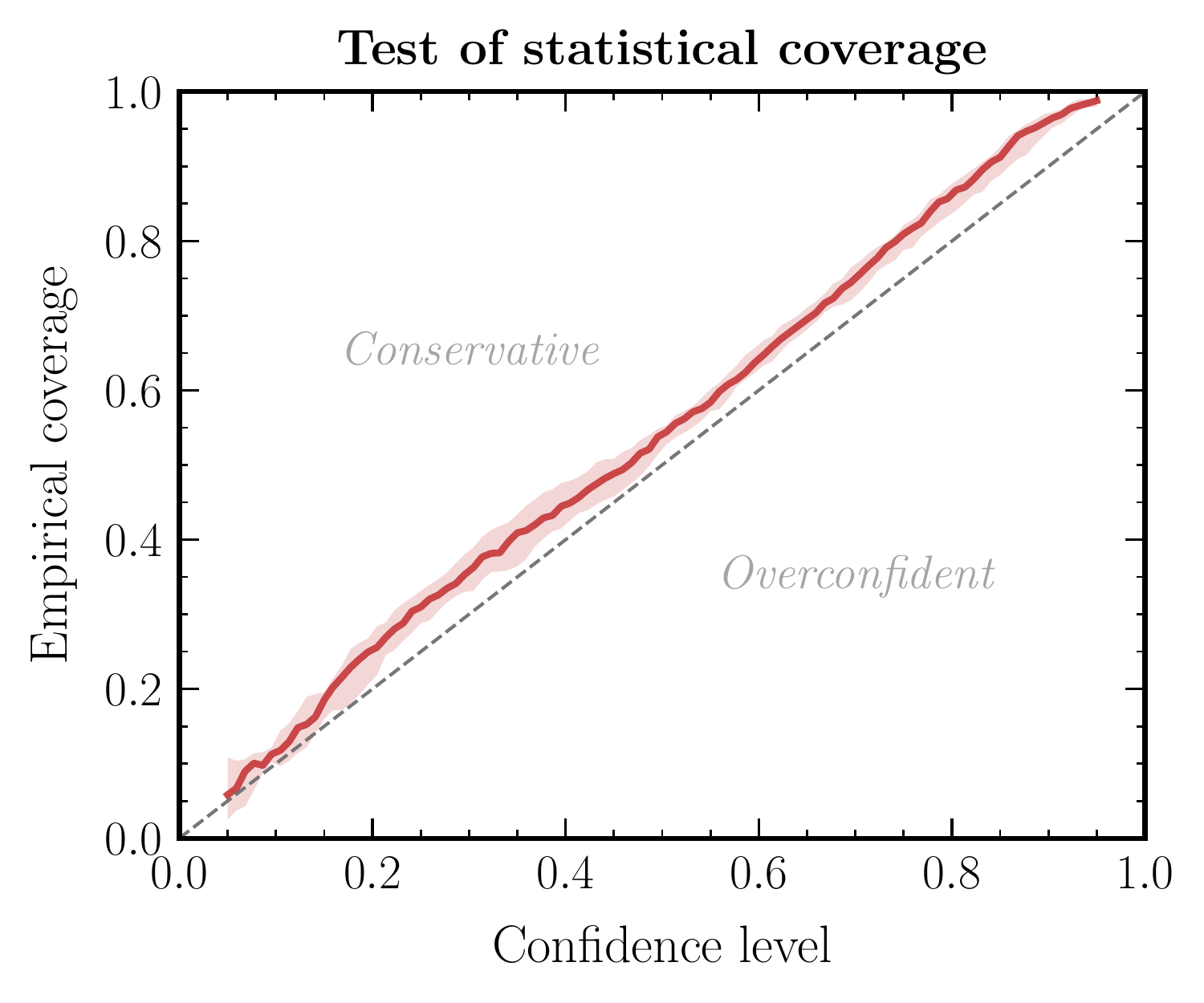}
\caption{{The empirical coverage of the baseline likelihood-ratio estimator as a function of nominal confidence level. The median (solid red line) and middle-68\% containment (red band) over 10 sets of 1000 samples is shown. The estimator is seen to have the desired property of being conservative, while still closely tracing the perfectly-calibrated regime (dashed grey line).}}
\label{fig:coverage}
\end{figure}

{Finally, we assess the quality of the approximate likelihood-ratio estimator through a test of statistical coverage. Within a hypothesis testing framework, this is necessary in order to ensure that the learned estimator is conservative over the parameter range of interest and does not produce overly confident or biased results~\citep{hermans2021averting}. We obtain the estimated TS profile for 1000 simulated samples with true substructure fraction values drawn from the range $f_\mathrm{sub}\in[0.1, 0.3]$. In doing so, we exclude parameter points towards the edges of our parameter space since the corresponding confidence intervals in these cases would extend outside of the tested parameter range, as can also be inferred from the baseline analysis shown in Fig.~\ref{fig:experiment}. For nominal confidence levels in the range $1-\alpha\in[0.05, 0.95]$ we compute the empirical coverage over the set of samples, defined as the fraction of samples whose true parameter values fall within the TS confidence interval. The confidence level for a given nominal confidence interval is computed under the assumption that the TS is $\chi^2$-distributed~\citep{10.1214/aoms/1177732360}. The procedure is repeated for 10 different sets of 1000 samples in order to estimate the statistical uncertainty associated with the empirical coverage.

The results of the coverage test are shown in Fig.~\ref{fig:coverage}, illustrating the median (solid red line) and middle-68\% containment (red band) of the empirical coverage. We see that the empirical coverage has the desired property of being conservative while still being close to the perfectly-calibrated regime (dashed grey line). We emphasize that this diagnostic tests the quality of the likelihood-ratio estimator over the entire evaluation parameter range $f_\mathrm{sub}\in[0.1, 0.3]$ rather than the baseline value $f_\mathrm{sub}\simeq0.2$ in isolation}.

\subsection{Experiments with unmodeled noise correlated on large scales}
\label{sec:lowell_noise}

\begin{figure*}[!htbp]
\centering
\includegraphics[width=0.95\textwidth]{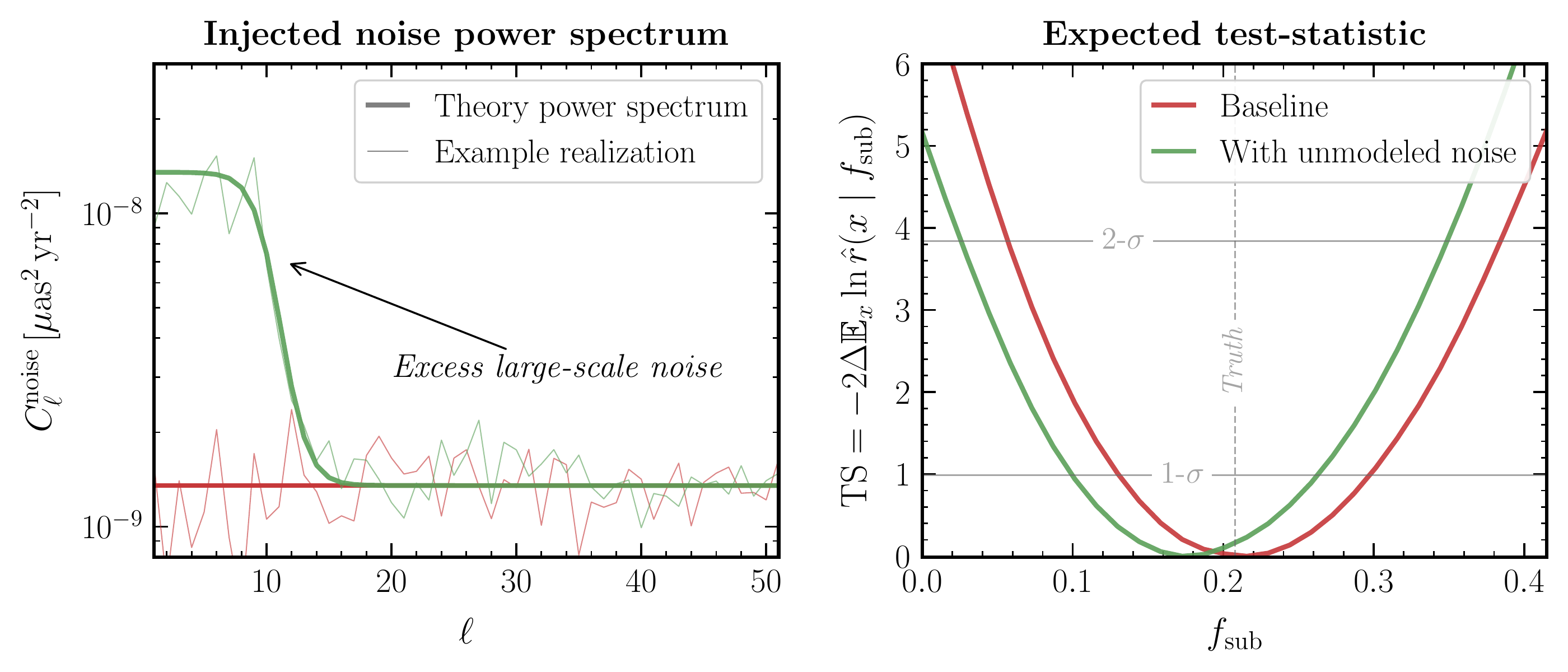}
\caption{\emph{(Left)} The power spectrum of the noise model (thicker green line) used to study the impact of correlated noise on large spatial scales, not modeled during training, on the performance of the likelihood-ratio estimator. The thinner green line shows the power spectrum of an example noise realization instantiated from this noise model. The red lines show corresponding power spectra for a scale-invariant noise model.  \emph{(Right)} The expected  test-statistic profile for a model evaluated on maps containing excess large-scale noise (green line) compared to the model evaluated on maps with scale-invariant noise (red line). A bias in the maximum-likelihood estimate returned by the model is seen when substantial unaccounted-for noise is presented in the test maps.}
\label{fig:noise_test}
\end{figure*}

Since the existence of measurement noise correlated on large spatial scales is a potential source of systematic uncertainty when working with astrometric maps, we test the susceptibility of our method to such effects by creating simulated data containing large-scale noise not previously seen by the trained estimator. Instead of assuming a scale-invariant noise power spectrum $C_\ell^\mathrm{noise} = 4\pi \sigma_{\mu}^2 / N_q$~\citep{Mishra-Sharma:2020ynk}, in this case we model noise with an order of magnitude excess in power on scales $\ell \lesssim 10$, parameterized as $C_\ell^\mathrm{noise} = 4\pi\sigma_{\mu}^2 / N_q \cdot \left(10 - 9 S(\ell - 10)\right)$ where $S$ denotes the sigmoid function.
The left panel of Fig.~\ref{fig:noise_test} illustrates this noise model (thicker green line) as well as the power spectrum of one simulated realization from this model (thinner green line, obtained using the \healpix module \texttt{anafast}) contrasted with the standard scale-invariant noise case (red lines). The right panel of Fig.~\ref{fig:noise_test} shows the expected log-likelihood ratio test-statistic profile for the two cases. Although a bias in the maximum-likelihood estimate of $f_\mathrm{sub}$ is seen when the test data has unmodeled noise (green line), the true test parameter value (dashed vertical line) is seen to lie well within the inferred 1-$\sigma$ confidence interval. This suggests that the method is only marginally susceptible to substantive amounts of correlated noise on large spatial scales.

\subsection{Experiments with a data-driven noise model}
\label{sec:dr2_noise}

{We finally assess the performance of our method using a realistic noise model obtained from the astrometric catalog of quasars in \emph{Gaia}'s Data Release 2 (DR2)~\citep{2018A&A...616A...1G,2018A&A...616A...2L}. The catalog contains the measured 2-dimensional positions, proper motions, as well as proper motion uncertainties of 555,934 quasars. Although the measured uncertainties in this case are too large for the catalog to be viable for the current scientific use-case, they can be rescaled and used to construct a data-driven noise model for testing the viability of our method on forthcoming astrometric data.

We compute the pixel-wise proper motion uncertainties as the inverse-variance weighted values within each \healpix pixel; $\sigma_{\mu}^\mathrm{pix} = \left(\sum_{q\in\mathrm{pix}}\sigma_{\mu, q}^{-2}\right)^{-1/2}$, where $\sigma_{\mu, q}^{2}$ are the provided variances of individual quasars within a given pixel. This results in a highly anisotropic noise model, shown in the left column of Fig.~\ref{fig:anisotropic_noise}, additionally having different uncertainties in the latitudinal (top row) and longitudinal (bottom row) directions. As expected due to occlusion from the Galactic disk, uncertainties are significantly higher towards the Galactic plane where the catalog has low completeness, additionally varying over the sky due to the scanning pattern and time-dependent instrumental response of the satellite. The region closest to the plane where no quasars are included in the catalog (shown in grey) is masked, testing the effect of partial sky coverage. In order to enable a direct comparison, the mean per-pixel variance for the data-driven noise model is normalized to that used in the baseline experiments in Sec.~\ref{sec:baseline}.

\begin{figure*}[!htbp]
\centering
\includegraphics[width=0.95\textwidth]{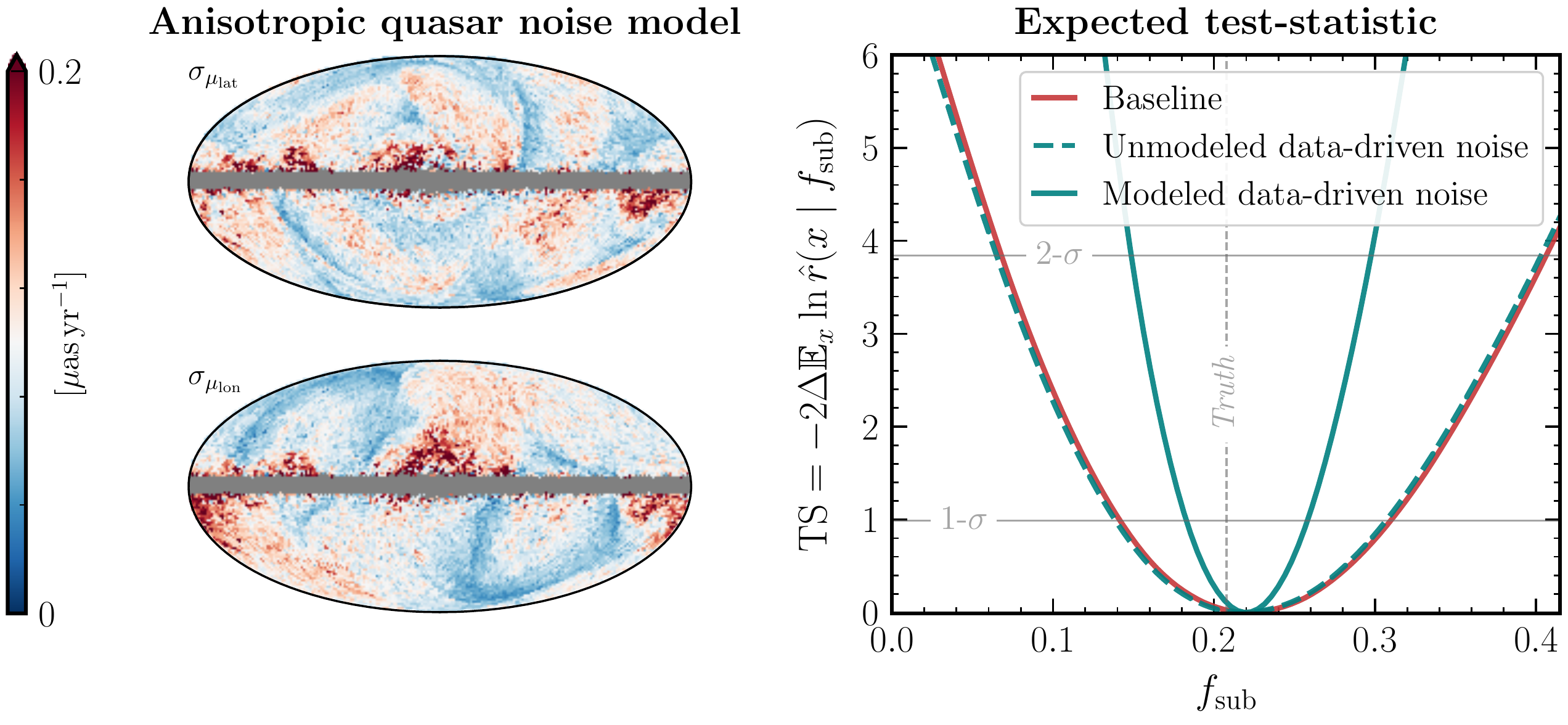}
\caption{{\emph{(Left)} The data-driven noise model derived using the \emph{Gaia} DR2 quasar catalog, showing maps of the effective per-quasar uncertainty within each pixel in the latitudinal (top) and longitudinal (bottom) directions. \emph{(Right)} The expected likelihood-ratio test statistic profile evaluated on samples with data-driven noise using the baseline estimator (dashed teal line) and the estimator trained with data-driven noise (solid teal line), compared with the baseline estimator evaluated on samples with isotropic noise (solid red line).}}
\label{fig:anisotropic_noise}
\end{figure*}

We evaluate the expected likelihood-ratio test statistic on samples generated with the data-driven noise model using two different estimators: \emph{(i)} the baseline estimator trained using samples with isotropic noise, as describe in Sec.~\ref{sec:baseline}, and \emph{(ii)} an estimator trained on samples generated with the correct, data-driven noise model used during evaluation. For \emph{(ii)}, a larger training batch size of 512 was found to provide better results, with all other hyperparameters being the same as the baseline case. The expected likelihood-ratio test statistic profiles for these cases are shown in the right column of Fig.~\ref{fig:anisotropic_noise}. Interestingly, evaluating the baseline estimator on samples with the data-driven noise model (dashed teal) produces results very similar to the baseline case (solid red). Using the correct, data-driven noise model on the other hand produces tighter constraints (solid teal) due to the fact that large portions of the sky in this case have smaller modeled uncertainties compared to the baseline case. In either case, successful recover of the astrometric lensing signal can be seen. These experiments demonstrate the viability of our method in the context of real-world applications, and we leave a more detailed study of our method in the context of forthcoming astrometric surveys and datasets to future work.}

\section{Conclusions and outlook}
\label{sec:conclusions}

We have introduced a method to analyze astrometric datasets over large regions of the sky using techniques based on machine learning with the aim of inferring the lensing signature of a dark matter substructure. We have shown our method to be significantly more sensitive to a cold dark matter subhalo population compared to established methods based on global summary statistics, with more favorable scaling as a function of measurement noise. Since the collection and reduction of astrometric data is an expensive endeavor, the use of methods that can take advantage of more of the available information can be equated to long periods of data-taking, underscoring their importance. Additionally, unlike the power spectrum approach, the current method does not require the construction of a numerically-expensive estimator to account for non-uniform exposure, selection effects, and instrumental noise in realistic datasets. These, as well as any other modeled observational effects, can be incorporated directly at the level of the forward model.

We have focused in this work on assessing sensitivity to a cold dark matter-like subhalo population with quasar velocity astrometry, which is within the scope of upcoming radio surveys like the SKA~\citep{Fomalont:2004hr,Jarvis:2015tqa}. Our method can also be applied in a straightforward manner to look for the \emph{acceleration} lensing signal imprinted on Milky Way stars, in particular sourced by a population of more compact subhalos than those expected in the cold dark matter scenario. These features are expected to imprint a larger degree of non-Gaussianity compared to the signal explored here (as can be seen, \eg, from Fig.~1 of \citet{Mishra-Sharma:2020ynk}), and machine learning methods may provide larger relative sensitivity gains when deployed in that context. Such analyses are within purview of the upcoming Roman exoplanet microlensing survey~\citep{Pardo:2021uzy} as well as future \emph{Gaia} data releases. {We note that when the region of interest covers a smaller fraction of the celestial sphere, as expected for the Roman microlensing survey of the Galactic bulge, the use of conventional convolutional architectures may be preferred as more efficient compared to spherical convolutions when the flat-sky approximation is valid.}

Several improvements and extensions to the method presented in this paper are possible. The use of architectures that can equivariantly handle vector inputs~\citep{DBLP:conf/nips/EstevesMD20} {may} aid in learning more efficient representations of the astrometric map. Using convolutions based on fixed rather than learned filters can additionally reduce model complexity and produce more interpretable representations~\citep{Cheng:2020qbx,2021arXiv210709145H,2021arXiv210411244S,2021arXiv210202828M,Valogiannis:2021chp}. The use of methods for likelihood-ratio estimation that can leverage additional latent information in the forward model can significantly enhance the sample efficiency of the analysis~\citep{Brehmer:2018eca,Brehmer:2018hga,Brehmer:2018kdj,Stoye:2018ovl}. We leave the study of these extensions as well as application of our method to other dark matter population scenarios to future work.

Astrometric lensing has been established as a promising way to characterize the Galactic dark matter population, with theoretical progress in recent years going in step with advances on the observational front. While this work is a first attempt at bringing principled machine learning techniques to this field, with the availability of increasingly complex datasets we expect machine learning to be an important general-purpose tool for future astrometric dark matter searches.

Code used for reproducing the results presented in this paper is available at \url{https://github.com/smsharma/neural-global-astrometry}. 

\begin{acknowledgments}
SM warmly thanks Kyle Cranmer, Cristina Mondino, Tess Smidt, Ken Van Tilburg, and Neal Weiner for helpful conversations. SM benefitted from the hospitality of the Center for Computational Astrophysics at the Flatiron Institute while this work was being performed. 
This work was performed in part at the Aspen Center for Physics, which is supported by National Science Foundation grant PHY-1607611.
The participation of SM at the Aspen Center for Physics was supported by the Simons Foundation.
This work is supported by the NSF CAREER grant PHY-1554858, NSF grants PHY-1620727 and PHY-1915409, and the Simons Foundation. 
This work is supported by the National Science Foundation under Cooperative Agreement PHY-2019786 (The NSF AI Institute for Artificial Intelligence and Fundamental Interactions, \url{http://iaifi.org/}).
This material is based upon work supported by the U.S. Department of Energy, Office of Science, Office of High Energy Physics of U.S. Department of Energy under grant Contract Number DE-SC0012567.
This work made use of the NYU IT High Performance Computing resources, services, and staff expertise. 
This research has made use of NASA's Astrophysics Data System. 
This work has made use of data from the European Space Agency (ESA) mission {\it Gaia} (\url{https://www.cosmos.esa.int/gaia}), processed by the {\it Gaia} Data Processing and Analysis Consortium (DPAC, \url{https://www.cosmos.esa.int/web/gaia/dpac/consortium}). Funding for the DPAC has been provided by national institutions, in particular the institutions participating in the {\it Gaia} Multilateral Agreement. 
We acknowledge the use of the \deepsphere graph convolutional layer implementation as well as code used to produce elements of Fig.~\ref{fig:model} from the code repository associated with \citet{2020arXiv201215000D} (\url{https://github.com/deepsphere/deepsphere-pytorch}).
\end{acknowledgments}

\software{
\package{Astropy}~\citep{Robitaille:2013mpa,Price-Whelan:2018hus},
\package{healpy}~\citep{Gorski:2004by,Zonca2019},
\package{IPython}~\citep{PER-GRA:2007},
\package{Jupyter}~\citep{Kluyver2016JupyterN},
\package{Matplotlib}~\citep{Hunter:2007},
\package{MLflow}~\citep{chen2020developments},
\package{NumPy}~\citep{harris_array_2020},
\package{PyGSP}~\citep{michael_defferrard_2017_1003158},
\package{PyTorch}~\citep{NEURIPS2019_9015},
\package{PyTorch Geometric}~\citep{Fey/Lenssen/2019}, 
\package{PyTorch Lightning}~\citep{william_falcon_2020_3828935},
\package{sbi}~\citep{tejero-cantero2020sbi},
\package{SciPy}~\citep{2020SciPy-NMeth}, and
\package{seaborn}~\citep{michael_waskom_2017_883859}.
}

\appendix

\section{{Additional details on the forward model}}
\label{app:forward_model_details}

We consider a population of Navarro-Frenk-White (NFW)~\citep{Navarro:1995iw} subhalos following a power-law mass function, $\mathrm dn / \mathrm dm \propto m^\alpha$, with slope $\alpha = -1.9$ as expected if the population is sourced from nearly scale-invariant primordial fluctuations in the canonical $\Lambda$ Cold Dark Matter ($\Lambda$CDM) scenario. The concentration-mass relation from \citet{Sanchez-Conde:2013yxa} is used to model the concentrations associated with density profiles of individual subhalos. 
Subhalos between $10^7$--$10^{10}\,\mathrm{M}_\odot$ are simulated, assuming the influence of lighter subhalos to be too small to be discernable~\citep{Mishra-Sharma:2020ynk}.

The spatial distribution of subhalos in the Galactocentric frame is modeled using results from the Aquarius simulation following \citet{Hutten:2016jko,Springel:2008cc}. Since this spatial distribution accounts for the depletion of subhalos towards the Galactic Center due to gravitational tidal effects, the angular number density of subhalos looking out from the Sun frame can be considered to be effectively isotropic.

The asymptotic velocities of subhalos in the Galactocentric frame are taken to follow a truncated Maxwell-Boltzmann distribution~\citep{1939isss.book.....C,Lisanti:2016jxe} $f_{\mathrm{Gal}}(\mathbf{v})\propto e^{-\mathbf{v}^{2} / v_{0}^{2}}\cdot H(v_\mathrm{esc} - |\mathbf{v}|)$, where $v_\mathrm{esc} = 550\,\mathrm{km}\,\mathrm{s}^{-1}$ is the Galactic escape velocity~\citep{Piffl:2013mla}, $v_\mathrm{0} = 220\,\mathrm{km}\,\mathrm{s}^{-1}$~\citep{Kerr:1986hz}, and $H$ is the Heaviside step function. Once instantiated, the positions and velocities of subhalos are transformed into the Galactic frame, assuming $R_\odot = 8.2\,\mathrm{kpc}$ to be the distance of the Sun from the Galactic Center~\citep{2019A&A...625L..10G,2020arXiv201202169B} and $\mathbf{v}_{\odot} = (11, 232, 7)\,\mathrm{km}\,\mathrm{s}^{-1}$ its Galactocentric velocity~\citep{2010MNRAS.403.1829S}. Note that the asymmetry in the direction of motion of the Sun in the Milky Way introduces a preferred direction for the Sun-frame velocities of subhalos, breaking strict rotation invariance in the forward model. Although not explicitly pursued here, {the expected characteristic form of this} asymmetry can be used as an additional distinguishing handle for the lensing signal, as was done in \citet{Mishra-Sharma:2020ynk}.

{Once a subhalo population has been instantiated using the forward model, the induced velocity lensing signal at different positions on the celestial sphere can be computed.} Given a spherically-symmetric subhalo lens moving with transverse velocity $\mathbf{v}_{l}$, the expected lens-induced velocity for a {background source} at impact parameter $\mathbf{b}$ is given by~\citep{VanTilburg:2018ykj}
\begin{equation}
\boldsymbol{\mu}(\mathbf{b})=4 G_{\mathrm{N}}\left\{\frac{M(b)}{b^{2}}\left[2 \hat{\mathbf{b}}\left(\hat{\mathbf{b}} \cdot \mathbf{v}_{l}\right)-\mathbf{v}_{l}\right]-\frac{M^{\prime}(b)}{b} \hat{\mathbf{b}}\left(\hat{\mathbf{b}} \cdot \mathbf{v}_{l}\right)\right\}
\end{equation}
where $M(b)$ and $M^{\prime}(b)$ are the projected mass of the subhalo at a given impact parameter distance $b = |\mathbf{b}|$ and its gradient. {In the context of our spatially-binned velocity map, $\mathbf{b}$ represents the vector from the center of the subhalo to the center of the respective \healpix pixel.}

\bibliography{astrometry-sbi}
\bibliographystyle{aasjournal-mod}

\end{document}